\documentclass[amsmath,amssymb,aps,twocolumn,showkeys,showpacs]{revtex4}
\usepackage[colorlinks=true, pdfstartview=FitV, linkcolor=blue, citecolor=red, urlcolor=magenta]{hyperref}
\usepackage{graphicx}
\usepackage{latexsym}
\usepackage{amsmath}
\usepackage{amsfonts}
\usepackage{amssymb}
\usepackage{verbatim}
\usepackage{dcolumn}
\usepackage{amssymb}
\usepackage{bm}
\usepackage{float}
\usepackage{subfigure}
\usepackage[english]{babel}
\newcommand{\be}{\begin{equation}}
\newcommand{\ee}{\end{equation}}
\newcommand{\bea}{\begin{eqnarray}}
\newcommand{\eea}{\end{eqnarray}}

\begin{document}

\newcommand{\NITK}{
\affiliation{Department of Physics, National Institute of Technology Karnataka, Surathkal, Mangaluru  575 025, India}
}

\newcommand{\IIT}{\affiliation{
Department of Physics, Indian Institute of Technology, Ropar, Rupnagar, Punjab 140 001, India
}}

\title{Photon Orbits and Thermodynamic Phase Transition of Regular AdS Black Holes}

\author{Naveena Kumara A.}
\email{naviphysics@gmail.com}
\NITK
\author{Ahmed Rizwan C.L.}
\email{ahmedrizwancl@gmail.com}
\NITK
\author{Shreyas Punacha}
\email{shreyasp444@gmail.com}
\NITK
\author{Md Sabir Ali}
\email{alimd.sabir3@gmail.com}
\IIT
\author{Ajith K.M.}
\email{ajith@nitk.ac.in}
\NITK

\begin{abstract}
We probe the phase structure of the regular AdS black holes using the null geodesics. The radius of photon orbit and minimum impact parameter shows a non-monotonous behaviour below the critical values of the temperature and the pressure, corresponding to the phase transition in extended phase space. The respective differences of the radius of unstable circular orbit and the minimum impact parameter can be seen as the order parameter for the small-large black hole phase transition, with a critical exponent $1/2$. Our study shows that there exists a close relationship between the gravity and thermodynamics for the regular AdS black holes. 

\end{abstract}


\maketitle

\section{Introduction}
The importance of black hole thermodynamics is indisputable due to its intriguing applications, since the seminal work of Hawking and Bekenstein \citep{Hawking:1974sw, Bekenstein1973}. The identification of temperature and entropy from the surface gravity and area of the event horizon, respectively, enable one to view the black hole as a thermodynamic system.  Interestingly, as in the case of a conventional thermodynamic system, black holes undergo phase transitions in a quite lot of ways. However, the phase transitions of black holes in AdS space have gained wide attention due to their thermal stability. In their pioneering work, Hawking and Page \citep{Hawking1983}  had shown the possibility of a phase transition between the thermal radiation and the large black hole in the AdS cavity. Since the introduction of proper pressure term using the cosmological constant in AdS space  \citep{Kastor:2009wy, Dolan:2011xt}, it was observed that a van der Waals (vdW) like phase transition is exhibited by the charged AdS black holes in the extended phase space \citep{Kubiznak2012, Gunasekaran2012, Kubiznak:2016qmn}. The phase transition in this scenario is between the small and large black holes (SBH-LBH), which is analogous to the liquid-gas transition in a vdW fluid.

The characteristic features of the material particle in the very vicinity of the event horizon can be utilized to unveil the information encoded in the concerned black hole. This motivation leads us to directly link the analysis of the particle motions, that are affected by the strong gravity near the compact objects such as a black hole, neutron stars etc., to the black hole properties. The study of geodesics of a test particle plays a vital role in the understanding of some observational effects such as the strong gravitational lensing and black hole silhouette, as well as quasinormal modes \citep{Cardoso:2008bp, Stefanov:2010xz}. Attempts to unravel the vdW phase transition of a black hole through astrophysical observations has its roots in quasinormal mode (QNM) studies \citep{Liu:2014gvf}. In these studies, it was reported that during the SBH-LBH phase transition, the slope of the quasinormal mode changes drastically. 

Prompted by the study relating the dynamics and thermodynamics, in the context of AdS black holes, recently there were attempts to establish a relationship between the gravity and thermodynamics \citep{Wei:2017mwc, Wei:2018aqm}. The correlation between the gravity and the critical behaviour is seen through the unstable null geodesic, which is encoded with the phase transition details. The radius of the photon sphere and the minimum impact parameter of the photon orbit exhibits an oscillatory behaviour during the vdW phase transition. Above the critical point of phase transition, the behaviour of these quantities become monotonous. This is analogous to the behaviour of Hawking temperature with horizon radius and entropy. Another significant result is that the respective differences in the radius and the minimum impact parameters act as an order parameter for SBH-LBH phase transition with a critical exponent $1/2$. The phase transition is scrutinised using photon orbit method for several black holes in different spacetime backgrounds \citep{Xu:2019yub, Chabab:2019kfs, Li:2019dai, Han:2018ooi, Hegde:2020yrd}. Studies related to null geodesics in other contexts have also appeared in subsequent works \citep{ Zhang:2019tzi, Bhamidipati:2018yqy,  Wei:2019jve}.

The Penrose censorship conjecture states the existence of singularity dressed by an event horizon \citep{Hawking:1969sw, Hawking:1973uf}. Therefore all the electrovacuum solutions of Einstein general relativity are in accordance with such point of view. However, such conjectures do not forbid us to consider the regular black hole spacetimes free from the singularity. Regular black holes were proposed to overcome such singular points, where the central singularity is replaced by a repulsive de-Sitter core. In this regard, motivated by the  ideas of Sakharov \citep{1966JETP...22..241S} and Gliner \citep{1966JETP...22..378G}, Bardeen proposed first regular black hole solution \citep{bardeen1968non}. The subsequent study of all the regular black holes was inspired by Bardeen's idea \citep{Hayward:2005gi, AyonBeato:1998ub, AyonBeato:2000zs}. Later Ayon-Beato-Garcia found the first exact regular black hole solution of Einstein field equations coupled to a nonlinear electrodynamic source. Various properties such as the black hole thermodynamics \citep{Man:2013hpa, Man:2013hza}, the rotating black hole shadows \citep{Abdujabbarov:2016hnw, Amir:2016cen}, quasinormal modes \citep{Flachi:2012nv} as well as the strong gravitational lensing \citep{Eiroa:2010wm} have been investigated in the background of regular black hole spacetimes. Some regular black hole solutions were also considered in alternative theories of gravity such as Lovelock gravity \citep{Aros:2019quj} and massive gravity theories \citep{Nam:2018ltb}. Regular black holes have also been extended to higher dimensions, to study its horizon structure and thermodynamical properties \citep{PhysRevD.98.084025,Kumar:2018vsm}. In our recent work, we have investigated the microstructure of the regular Hayward black hole using the Ruppeiner geometry method, where we have reported the existence of repulsive interaction in the black hole microstructure \citep{Kumara:2020mvo, Kumara:2020ucr}. In the present work, we study the phase transition of regular black holes in AdS spacetime by considering the correspondence between photon orbits and the extended phase space thermodynamics. We show the parametric effect induced in the black hole solution due to the presence of nonlinear charge.

The organisation of the paper is as follows. In the next section (\ref{regularsection}), we present the construction of regular black hole solutions in general relativity. In section \ref{haywardsection}, we discuss the phase transition of regular Hayward AdS black hole using the photon orbits. In section  \ref{bardeensecion} we carry out a similar investigation for the regular Bardeen AdS black hole. Finally, we conclude the paper in section \ref{conclusion}.

\section{Construction of Regular Black Hole Solutions}
\label{regularsection}

We first present the regular black holes solutions in the background of anti-de Sitter spacetime. The solutions we are interested in, the Hayward and the Bardeen black holes, can be derived from Einstein gravity minimally coupled to nonlinear electrodynamics with negative cosmological constant $\Lambda$ given by the action \cite{Fan:2016hvf},

\begin{equation}
\label{action}
    \mathcal{I}=\frac{1}{16\pi G} \int{d^4x}\sqrt{-\hat{g}}[R-\mathcal{L\left(F\right)}+2\Lambda],
\end{equation}
where $R$ and  $\hat{g}$ are the Ricci scalar and the determinant of the metric tensor, respectively. $\mathcal{L(F)}$ is the Lagrangian density of nonlinear electrodynamics which is a function of $\mathcal{F} = F_{\mu \nu }F^{\mu \nu }$ with $F_{\mu\nu}=2\nabla_{[\mu}A_{\nu]}$, the strength tensor of nonlinear electrodynamics. Varying the action (Eq. {\ref{action}}), with respect to $g_{\mu\nu}$ and $A_{\mu}$, we have the field equations of the form,
\begin{eqnarray}\label{fieldeqs1}
G_{\mu\nu}+\Lambda g_{\mu\nu}&=&T_{\mu\nu},
\end{eqnarray}
\begin{eqnarray}\label{fieldeqs2}
\nabla_{\mu}\left(\frac{\partial \mathcal{L\left(F\right)}}{\partial{F}}{F^{\nu\mu}}\right)&=&0,\qquad \nabla_{\mu}\left(*F^{\nu\mu}\right)=0,
\end{eqnarray}
where $T_{\mu\nu}$ is energy-momentum tensor, which can be written as
\begin{eqnarray}
T_{\mu\nu}=2\left[\mathcal{L}_{\mathcal{F}}F_{\mu\alpha}F_{\nu}^{\alpha}-\frac{1}{4} g_{\mu\nu}\mathcal{L\left(F\right)}\right],
\label{EMT8}
\end{eqnarray}
where $\mathcal{L}_{\mathcal{F}}=\frac{\partial\mathcal{L\left(F\right)}}{\partial{F}}$. In this article, we contemplate static spherically symmetric black holes with magnetic charges. To construct such black hole solutions, we follow the general procedure given as in Ref. \cite{Fan:2016hvf}. The regular Hayward black hole solution can be obtained from the Lagrangian density,
\begin{eqnarray}
\label{lagran}
\mathcal{L\left(F\right)}=\frac{12}{\alpha}\frac{\left(\alpha \mathcal{F}\right)^{3/2}}{\left(1+\left(\alpha \mathcal{F}\right)^{3/4}\right)^2},
\end{eqnarray}
with $\alpha >0$ which has the dimension of length squared. The four-dimensional spherically symmetric black hole is described by 
\begin{eqnarray}\label{metric}
ds^2&=&-\left(1-\frac{2m(r)}{r}\right)dt^2+\frac{dr^2}{\left(1-\frac{2m(r)}{r}\right)}+r^2d\Omega_{2}^2,
\end{eqnarray}
where $m(r)$ is the mass function containing the mass within radius $r$ and $d\Omega_{2}^2=d\theta^2+\sin^2\theta{d\phi^2}$, is a $2$-dimensional unit sphere. For a spherically symmetric spacetime, $F_{\mu\nu}$ admits two non-vanishing components, $F_{tr}$ and $F_{\theta\phi}$. For a pure magnetic charge only $F_{\theta\phi}$ survives. The ansatz for $F_{\mu\nu}$ for a purely magnetically charged black hole reads, 
\begin{eqnarray}\label{Fmunu}
F_{\mu\nu}=2\delta^{\theta}_{[\mu}\delta^{\phi}_{\nu]}\mathcal{X}\left(r,\theta\right).
\end{eqnarray}
Utilizing Eq.~(\ref{Fmunu}) in Eq.~(\ref{fieldeqs2}) and integrating it, we have,
\begin{eqnarray}\label{Fmunu1}
F_{\mu\nu}=2\delta^{\theta}_{[\mu}\delta^{\phi}_{\nu]}q\left(r\right)\sin\theta.
\end{eqnarray}
Eq.~(\ref{fieldeqs2}) implies $dF=0$ which in turn reads $q^\prime(r)dr\wedge d\theta\wedge d\phi=0$, leading to $q(r)=constant=Q_m$. The constant $Q_m$ is identified with the magnetic monopole charge of the nonlinear electrodynamics. Then, the resulting Maxwell tensor reads with only component,
\begin{eqnarray}\label{maxwell}
F_{\theta\phi}=-F_{\phi\theta}=-Q_m\sin\theta,
\end{eqnarray}
and hence the gauge potential and the Maxwell invariant for this field turn out to be,
\begin{eqnarray}
\label{gauge}
A_\mu=Q_m\cos\theta\delta_\mu^\phi, \qquad \mathcal{F}=\frac{2 Q_m^2}{r^4}.
\end{eqnarray}
Using this result the Lagrangian (\ref{lagran}) can be casted as,
\begin{eqnarray}\label{lagran1}
\mathcal{L}\left(r\right)=\frac{12}{\alpha} \frac{\left(2\alpha Q_m^2\right)^{3/2}}{\left(r^3+(2\alpha Q_m^2)^{3/4}\right)^2}.
\end{eqnarray}
The two independent non-zero components of the Einstein field equations, using energy-momentum tensor (Eq.~\ref{EMT8}), can be obtained as
\begin{eqnarray}
\label{emt-1}
\frac{2m^\prime(r)}{r^2}-\Lambda=\mathcal{L}(r),\\
\frac{m^{\prime\prime}(r)}{r}-\Lambda=\left(\mathcal{L}(r)-\mathcal{L}_{\mathcal{F}}(r)F^{\theta\phi}F_{\theta\phi}\right).
\end{eqnarray}
The solution of Eq.(\ref{emt-1}) is calculated to be,
\begin{eqnarray}\label{massf1}
m(r)=\frac{2 M r^2}{r^3+g^3}+\frac{\Lambda r^3}{6},
\end{eqnarray}
where $M$ the mass of the black hole and $g$ the free integration constant that is related to the magnetic charge $Q_m$, are identified as,
\begin{equation}
M=\alpha^{-1}{g^3}, \qquad Q_m=\frac{g^2}{\sqrt{2\alpha}}.
\end{equation}
Thus, from Eq.~(\ref{metric}) the metric for a regular Hayward AdS black hole in four-dimensional spacetime reads,
\begin{eqnarray}\label{metric2}
\mathrm{ds^2}=-f(r)dt^2+\frac{1}{f(r)}dr^2+r^2 d\Omega_{2}^2,
\end{eqnarray}
with the metric function
\begin{equation}
    f(r)=\left(1-\frac{2 M r^2}{r^3+g^3}-\frac{\Lambda r^2}{3}\right).
\end{equation}

Similar procedures follow for the regular Bardeen AdS black hole with the following Lagrangian density,

\begin{eqnarray}
\label{lagranb}
\mathcal{L\left(F\right)}=\frac{12}{\alpha}\frac{\left(\alpha \mathcal{F}\right)^{5/4}}{\left(1+\left(\alpha \mathcal{F}\right)^{1/2}\right)^{5/2}}.
\end{eqnarray}
The corresponding line element for a spherically symmetric spacetime has the same form as Eq. (\ref{metric2}), with the metric function,
\begin{equation}
 f(r)=   \left(1-\frac{2 M r^2}{(r^2+g^2)^{3/2}}-\frac{\Lambda r^2}{3}\right).
\end{equation}

\section{Regular Hayward Black Hole}
\label{haywardsection}
\subsection{Thermodynamics of Regular Hayward Black Hole}
Here we briefly present the extended thermodynamics of the regular Hayward black hole. In extended phase space description, the pressure $P$ is related to the cosmological constant $\Lambda$ as \citep{Kastor:2009wy, Dolan:2011xt},
\begin{equation}
   P= -\frac{\Lambda}{8\pi}.  
\end{equation}
The metric function thus takes the form,
\begin{equation}
f(r)=1-\frac{2 M r^2}{g^3+r^3}+\frac{8}{3} \pi  P r^2.
\end{equation}
The horizon of the black hole is characterised by the condition, $f(r_+)=0$. Using this condition we get,
\begin{equation}
    M=\frac{r_+}{2}+\frac{4}{3} \pi   P (g^3+r_+^3)+\frac{g^3}{2 r_+^2}.
\end{equation}
The Hawking temperature of the black hole which is associated to the surface gravity $\kappa$ is obtained as,
\begin{align}
T&=\frac{\kappa}{2\pi}=\left. \frac{f'(r)}{4\pi} \right|_{r=r_+}\nonumber \\
&=\frac{2 P r_+^4}{g^3+r_+^3}-\frac{g^3}{2 \pi  r_+ \left(g^3+r_+^3\right)}+\frac{r_+^2}{4 \pi  \left(g^3+r_+^3\right)}.
\end{align}
With these, the first law of thermodynamics reads as,
\begin{equation}
dM=TdS+\Psi dQ_m+VdP+\Pi d \alpha,
\end{equation}
where $\Psi$ and $\Pi$ are the variables conjugate to the magnetic charge $Q_m$ and parameter $\alpha$, respectively. The entropy  and volume of the black hole have the following non trivial profile,
\begin{align}
S&=\int \frac{dM}{T}=2 \pi  \left(\frac{r_+^2}{2}-\frac{g^3}{r_+}\right),\\
V&=\left( \frac{\partial M}{\partial P}\right)_{S,Q_m,\alpha}=\frac{4}{3} \pi  \left(g^3+r_+^3\right).
\end{align}
We also note that the definition of entropy $S$ for regular black hole is not unique, another choice of $S$ from area law is also possible. Then we have to give up with the above first law, which is to be modified with a modified mass $M$ \citep{Ma:2014qma}. However these do not alter the phase transition and related properties of the black hole. The equation of state reads as,
\begin{equation}
P=\frac{g^3}{4 \pi  r_+^5}+\frac{g^3 T}{2 r_+^4}-\frac{1}{8 \pi  r_+^2}+\frac{T}{2 r_+}.
\end{equation}
The regular black hole exhibits a vdW like critical behaviour, which has been studied extensively in the literature \citep{Fan:2016rih}. The first order phase transition takes place between a small black hole (SBH) phase and a large black hole (LBH) phase. The critical point of this phase transition is given by,
\begin{equation}
T_{cH}=\frac{\left(5 \sqrt{2}-4 \sqrt{3}\right) \left(3 \sqrt{6}+7\right)^{2/3}}{4\times 2^{5/6} \pi  g},
\end{equation}
\begin{equation}
P_{cH}=\frac{3 \left(\sqrt{6}+3\right)}{16\times 2^{2/3} \left(3 \sqrt{6}+7\right)^{5/3} \pi  g^2},
\end{equation}
\begin{equation}
S_{cH}=\left(6 \sqrt{6}+14\right)^{2/3} \pi  g^2.
\end{equation}
It is clear that the critical values of the thermodynamic variables depend on the parameter $g$. Using this, the reduced thermodynamic variables are defined as,
\begin{equation}
    \tilde{T}=\frac{T}{T_{cH}} \qquad \tilde{P}=\frac{P}{P_{cH}} \qquad \tilde{V}=\frac{V}{V_{cH}}.
\end{equation}

\subsection{Geodesic equations of motion}
To obtain the relationship between the null geodesics and the phase transition of the black hole we consider a free photon orbiting around the black hole on the equatorial plane, i.e., $\theta =\pi /2$. Then the Lagrangian is,
\begin{equation}
2 \mathcal{L}=-f(r)\dot{t}^2+\frac{\dot{r}^2}{f(r)}+r^2\dot{\phi}^2.
\end{equation}
The dots over variables stand for the differentiation with respect to an affine parameter. The generalised momenta corresponding to this Lagrangian can easily be obtained as,
\begin{eqnarray}
p_t=-f(r)\dot{t}\equiv E\\
p_\phi= r^2\dot{\phi} \equiv L\\
p_r=\dot{r}/f(r).
\end{eqnarray}
In the above, $E$ and $L$ are the energy and orbital angular momentum of the photon, respectively, which are the constants of motion. The $t$ motion and $\phi$ motion can be written as,
\begin{equation}
\dot{t}=\frac{E}{f(r)}
\end{equation}
\begin{equation}
\dot{\phi}=\frac{L}{r^2 \sin ^2\theta}.
\end{equation}
The Hamiltonian for the system is,
\begin{equation}
2\mathcal{H}=-E\dot{t}+L\dot{\phi}+\dot{r}^2/f(r)=0.
\end{equation}

\begin{figure}[t]
\centering
\includegraphics[scale=0.9]{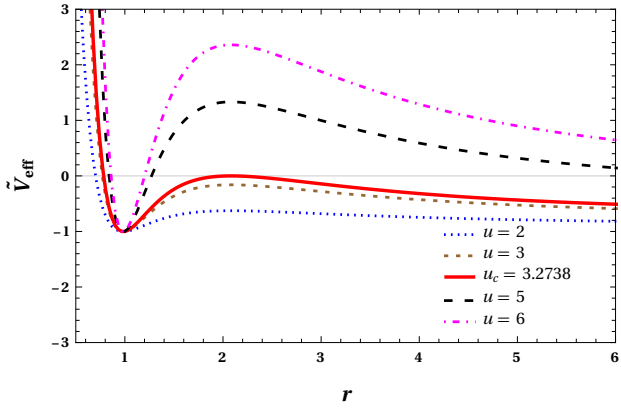}
\caption{The effective potential for the regular Hayward black hole. Here we take black hole horizon $r_+=1$, charge parameter $g=0.8$ and pressure  $P=0.003$. The thick red line corresponds to the critical angular momentum $L_c$ (corresponding critical impact parameter $u_c$)}
\label{veffHayward}
\end{figure}
The expression for the radial $r$ motion is rewritten as, 
\begin{equation}
\dot{r}^2+V_{eff}=0
\end{equation}
where $V_{eff}$ is the effective potential, which has the following explicit form,
\begin{equation}
V_{eff}=\frac{L^2}{r^2}f(r)-E^2.
\end{equation}
The behaviour of $\tilde{V}_{eff}=V_{eff}/E^2$ is shown in Fig. \ref{veffHayward} for different values of impact parameter $u=L/E$.

The accessible region for the photon is $V_{eff}<0$, since $\dot{r}^2>0$. From the Fig. \ref{veffHayward} it is clear that, the photon fall into the black hole for small values of $u$, whereas it is reflected for large values of $u$, as it approaches the black hole. Between these two conditions there is an unstable circular photon orbit which corresponds to the critical angular momentum (red thick line in Fig. \ref{veffHayward}). At the peak of that particular effective potential the radial velocity of the photon is zero. The corresponding value of $r$ at the peak is the radius of the photon sphere. The unstable circular orbit is characterised by,
\begin{equation}
V_{eff}=0\quad , \quad V'_{eff}=0 \quad , \quad V''_{eff}<0,
\end{equation}
where the prime denotes the differentiation with respect to $r$. Expanding the second equation $(V'_{eff}=0)$,
\begin{equation}
2f(r_{ps})-r_{ps}\partial _r f(r_{ps})=0.
\label{aneqn}
\end{equation}
The solution of this gives the radius of photon sphere $r_{ps}$,
\begin{widetext}
\begin{equation}
r_{ps}=\frac{1}{4} \left(2 \sqrt{-\frac{8 g^3 M}{\sqrt[3]{Y}}+\frac{9 M^2}{2}+\frac{27 M^3}{4 \sqrt{\frac{9 M^2}{4}+X+\sqrt[3]{Y}}}-\sqrt[3]{Y}}+\sqrt{9 M^2+4 \left(X+\sqrt[3]{Y}\right)}+3 M\right),
\label{rpsH}
\end{equation}
\end{widetext}
where,
\begin{equation}
X=\frac{8 g^3 M}{\sqrt[3]{27 g^3 M^3+\sqrt{729 g^6 M^6-512 g^9 M^3}}},
\end{equation}
and 
\begin{equation}
Y=27 g^3 M^3+\sqrt{729 g^6 M^6-512 g^9 M^3}.
\end{equation}

\begin{figure*}[t]
\centering
\subfigure[ref1][\, $\tilde{T}$ vs. $\tilde{r}_{ps}$]{\includegraphics[scale=0.9]{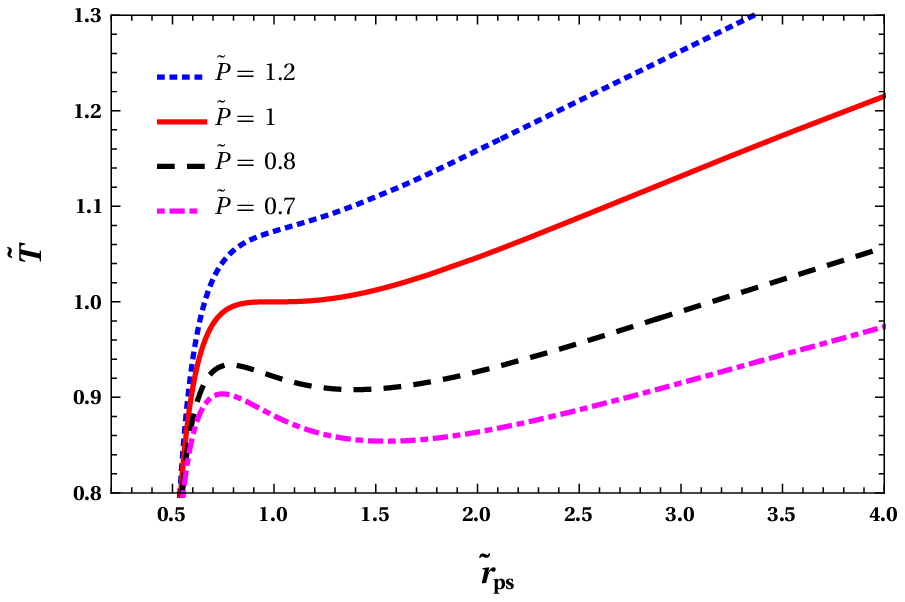}\label{TrHayward}}
\qquad
\subfigure[ref2][\, $\tilde{T}$ vs. $\tilde{u}_{ps}$]{\includegraphics[scale=0.9]{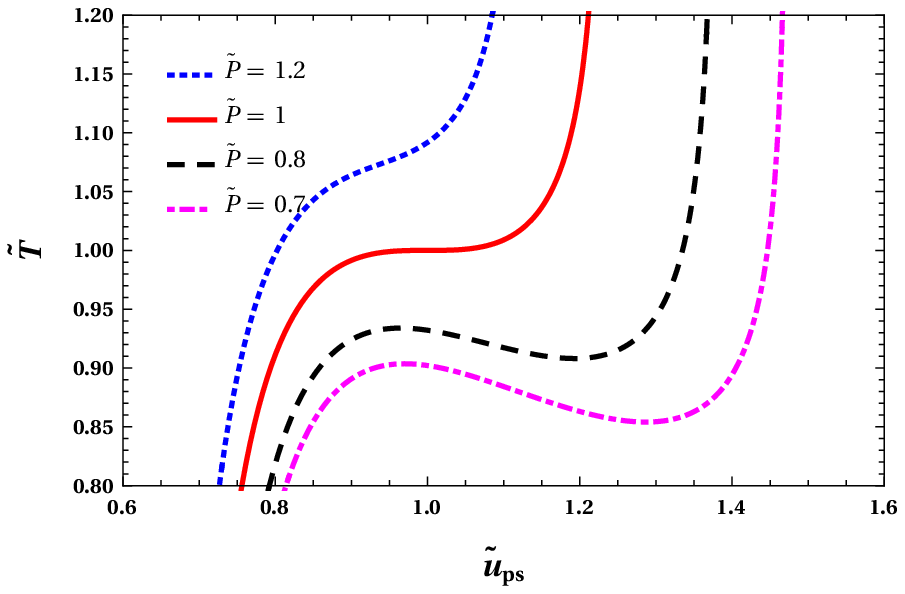}\label{TuHayward}}
\caption{The behaviour of photon sphere radius $r_{ps}$ and minimum impact parameter $u_{ps}$ with temperature in reduced space for Hayward case. These plots are for a fixed value of $g=0.8$ and the reduced pressure $\tilde{P}=0.7, 0.8, 1, 1.2$}
\label{TruHayward}
\end{figure*}

\begin{figure*}[t]
\centering
\subfigure[ref1][\, $\tilde{P}$ vs. $\tilde{r}_{ps}$]{\includegraphics[scale=0.9]{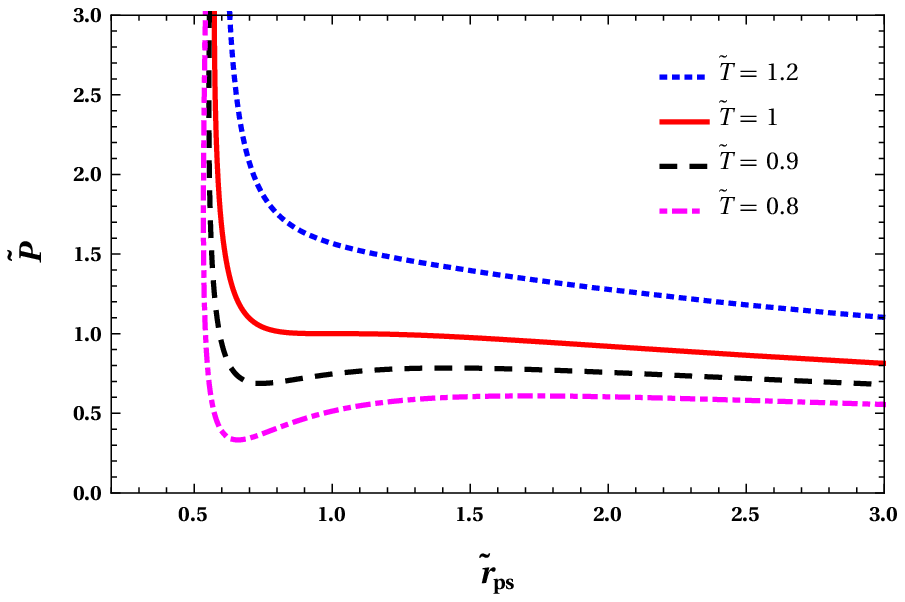}\label{PrHayward}}
\qquad
\subfigure[ref2][\, $\tilde{P}$ vs. $\tilde{u}_{ps}$]{\includegraphics[scale=0.9]{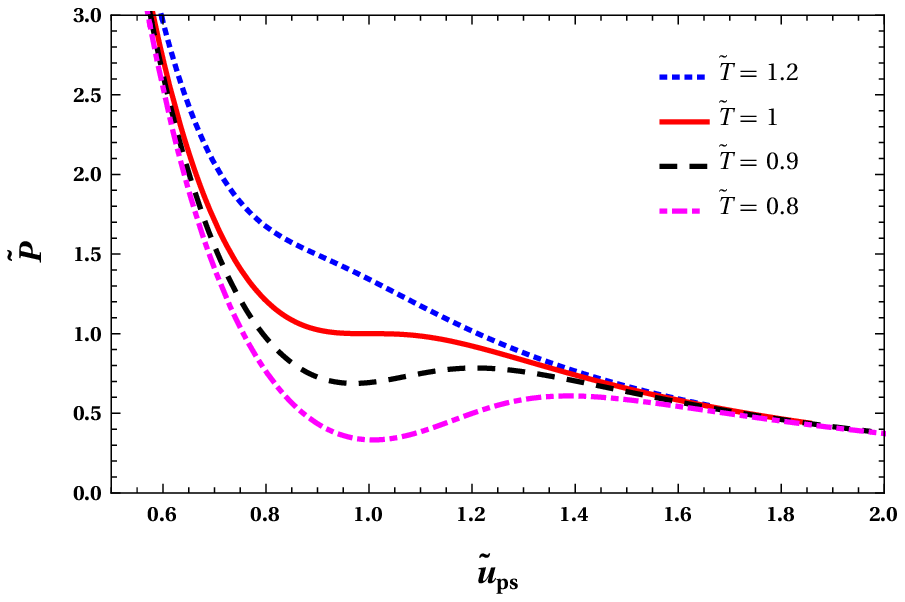}\label{PuHayward}}
\caption{The behaviour of photon sphere radius $r_{ps}$ and minimum impact parameter $u_{ps}$ with pressure in reduced space for Hayward case. These plots are for a fixed value of $g=0.8$ and the reduced temperature $\tilde{T}=0.8, 0.9, 1, 1.2$}
\label{PruHayward}
\end{figure*}

The solution of the first equation, $(V_{eff}=0)$, gives the minimum impact parameter of the photon,
\begin{equation}
u_{ps}=\frac{L_c}{E}=\left. \frac{r}{\sqrt{f(r)}} \right| _{r_{ps}}.
\label{upsequation}
\end{equation}
The explicit form of this can be obtained by using Eq. (\ref{rpsH}). 

The insight of using the photon sphere parameters $r_{ps}$ and $u_{ps}$ to probe the details of the small-large black hole phase transition, stems from the phenomenon of black hole lensing \cite{Wei:2017mwc}. In black hole lensing, the impact parameter of the photon $u$ has a close connection with the deflection angle. Larger the impact parameter smaller is the deflection angle. However, under the limit $u\rightarrow u_{ps}$ the deflection angle is unbounded \cite{Bozza:2002zj}. We can relate these key quantities, $r_{ps}$ and $u_{ps}$, to the thermodynamic variables $P$ and $S$ by using the expression for mass of the black hole $M$. $r_{ps}(P,S)$ is a complicated expression which we have not written here. The behaviour $r_{ps}$ and $u_{ps}$ against the temperature is studied in reduced parameter space, Fig. \ref{TrHayward} and \ref{TuHayward}. The similar study is carried out for $\tilde{P}-\tilde{r}_{ps}$ and $\tilde{P}-\tilde{u}_{ps}$ plots for a fixed value of reduced temperature, Fig. \ref{PrHayward} and \ref{PuHayward}.

In Fig. \ref{TruHayward}, both the photon sphere radius and the critical impact parameter shows non-monotonous behaviour below the critical pressure. All the isobars for $\tilde{P}<1$ have one minimum and a maximum. There is an inflexion point for the isobar $\tilde{P}=1$. For all the values above $\tilde{P}=1$, the oscillating behaviour disappears. This behaviour of $\tilde{T}-\tilde{r}_{ps}$ and $\tilde{T}-\tilde{u}_{ps}$ is quite similar to the isobars in $\tilde{T}-\tilde{S}$ plane for the vdW fluid. On the other hand the $\tilde{P}-\tilde{r}_{ps}$ and $\tilde{P}-\tilde{u}_{ps}$ plots  shown in Fig. \ref{PruHayward} have similarity with the isotherms of the vdW system in the $P-V$ plane. These connections indicate that there is a relationship between the photon orbits and the critical behaviour of the back hole.

\begin{figure*}[t]
\centering
\subfigure[ref1][\, $\tilde{r}_{ps}$ vs. $\tilde{T}$]{\includegraphics[scale=0.9]{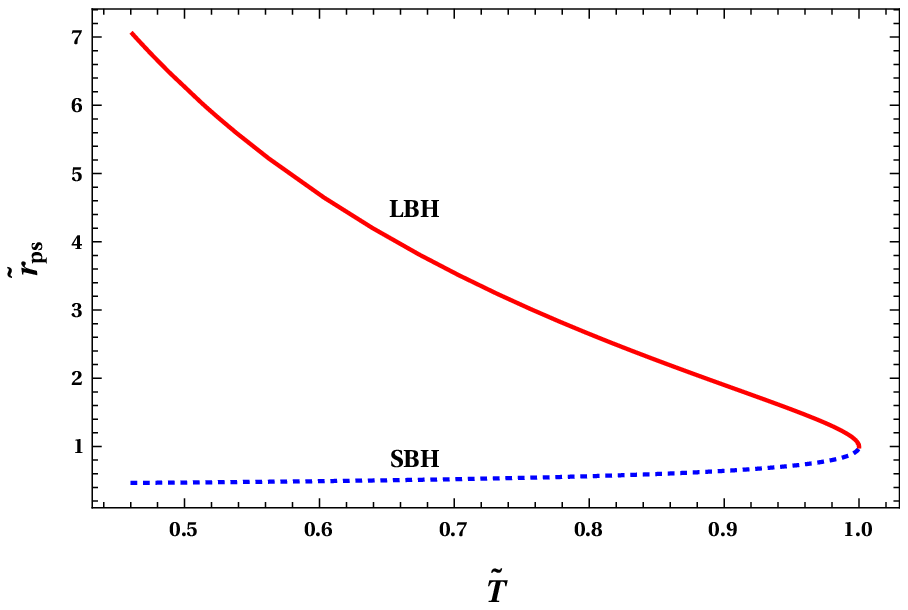}\label{rHayward}}
\qquad
\subfigure[ref1][\, $\tilde{u}_{ps}$ vs. $\tilde{T}$]{\includegraphics[scale=0.9]{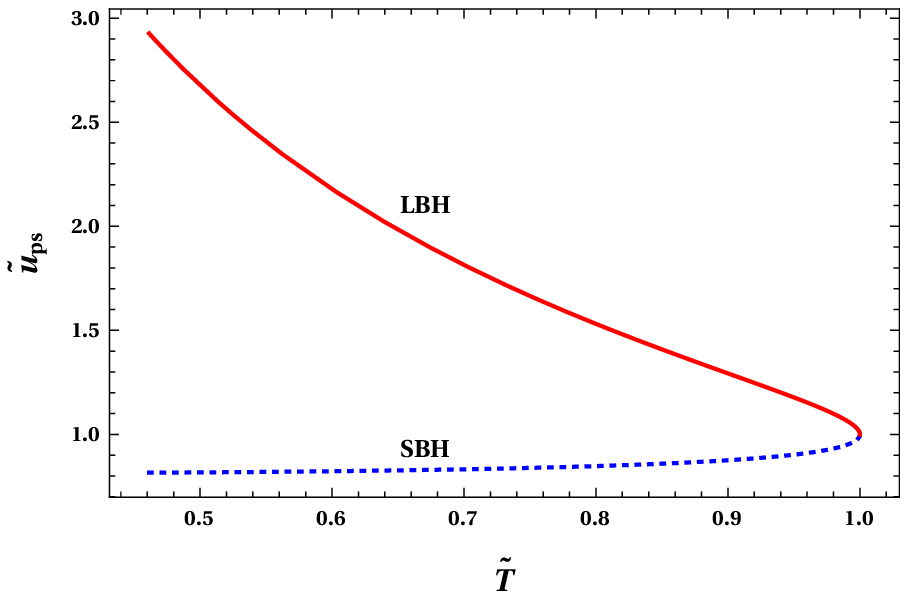}\label{uHayward}}
\caption{The behaviour of photon sphere radius $r_{ps}$ and minimum impact parameter $u_{ps}$ of the photon orbit along the coexistence curve for Hayward case}
\label{coex}
\end{figure*}

\begin{figure*}[t]
\centering
\subfigure[ref2][\,  $\Delta \tilde{r}_{ps}$ vs. $\tilde{T}$]{\includegraphics[scale=0.9]{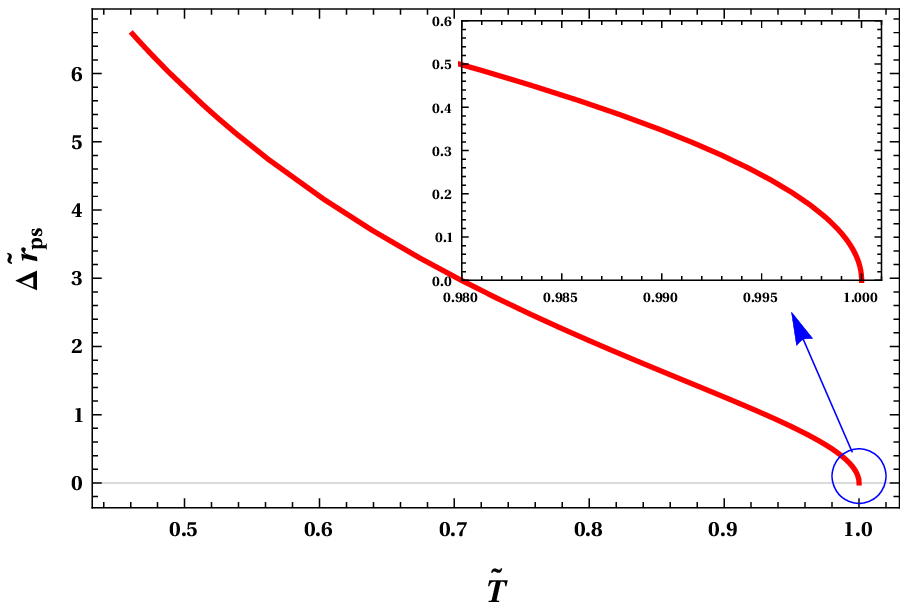}\label{drHayward}}
\quad
\subfigure[ref2][\, $\Delta \tilde{u}_{ps}$ vs. $\tilde{T}$]{\includegraphics[scale=0.9]{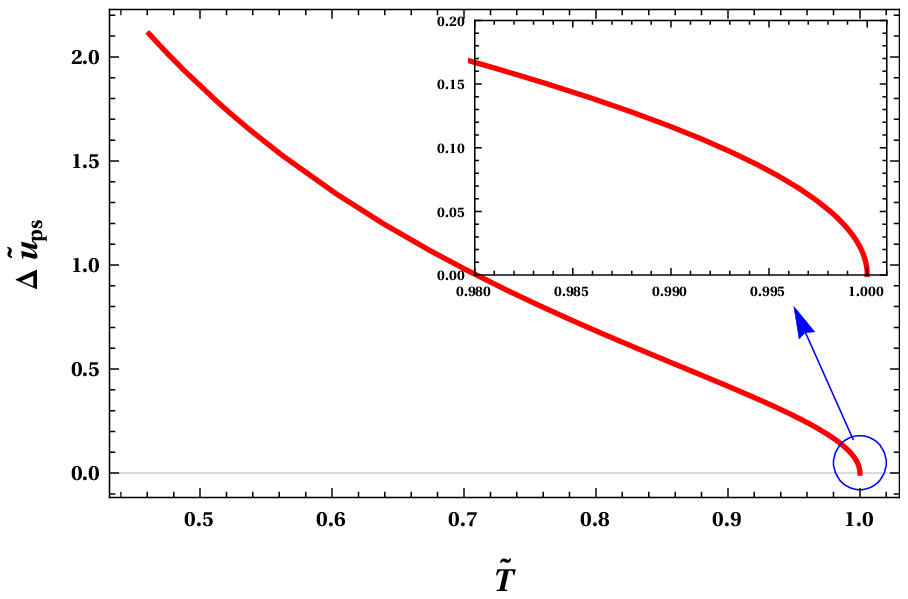}\label{duHayward}}
\caption{ The behaviour of difference of the radii of the circular orbit $\Delta \tilde{r}_{ps}$ and the difference of the minimum impact parameter $\Delta \tilde{u}_{ps}$ along the coexistence curve for Hayward case. The change of concavity near the critical point is shown in the inlets}
\label{diffcoex}
\end{figure*}

\subsection{Critical behaviour from unstable photon orbits}
We have constructed the equal area law for the  $\tilde{T}-\tilde{r}_{ps}$ and $\tilde{T}-\tilde{u}_{ps}$ isobars, similar to the isobars in the $\tilde{T}-\tilde{S}$ plane of vdW system. Using those results, we obtained the behaviour of the radius of the circular orbit and the minimum impact parameter along the coexistence curve (Fig. \ref{coex}). The $\tilde{r}_{ps}$ and $\tilde{u}_{ps}$ has two branches corresponding to SBH and LBH phases of the black hole. With the increase in temperature, both the $r_{ps}$ and $u_{ps}$ decreases for the LBH branch, whereas increases for SBH branch. At the critical value $\tilde{T}=1$ both branches share the same value. The differences $\Delta \tilde{r}_{ps}$ and $\Delta \tilde{u}_{ps}$ are plotted against the reduced temperature $\tilde{T}$, which is shown in Fig. \ref{diffcoex}. A sudden change in $\Delta \tilde{r}_{ps}$ and $\Delta \tilde{u}_{ps}$ exists in the regions corresponding to the first order phase transitions, i.e., for $\tilde{T}<1$. The difference becomes zero as the critical value of $\tilde{T}$ is approached, where the second order phase transition is observed.  The behaviour of $\Delta \tilde{r}_{ps}$ and $\Delta \tilde{u}_{ps}$ near the critical point is observed in the inlets of Fig. \ref{drHayward} and \ref{duHayward}. We also note that, near the critical point the concavity of the curve changes. The behaviour near the critical point can be assumed to be of the following form,
\begin{equation}
    \Delta \tilde{r}_{ps}, \quad  \Delta \tilde{u}_{ps} \sim a\times (1-\tilde{T})^\delta.
    \label{logfit}
\end{equation}
Taking logarithm on both side, we have,
\begin{equation}
    \ln  \Delta \tilde{r}_{ps} , \quad \ln \Delta \tilde{u}_{ps} \sim \delta \ln (1-\tilde{T})+\ln a.
\end{equation}

\begin{figure*}[t]
\centering
\subfigure[][\,  $ \ln \Delta \tilde{r}_{ps}$ vs. $\ln (1-\tilde{T})$]{\includegraphics[scale=0.9]{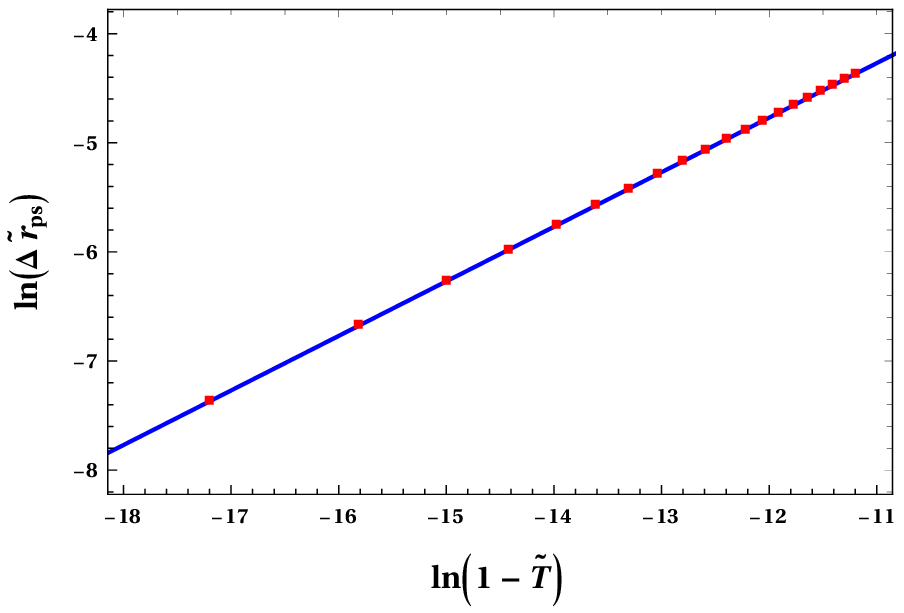}\label{drfitHayward}}
\quad
\subfigure[][\, $\ln \Delta \tilde{u}_{ps}$ vs. $\ln (1-\tilde{T})$]{\includegraphics[scale=0.9]{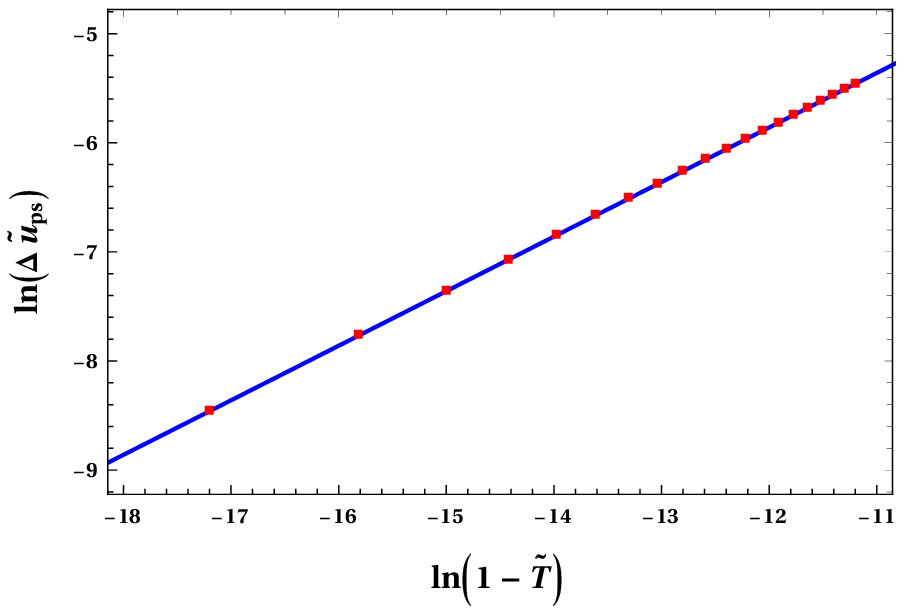}\label{dufitHayward}}
\caption{Near critical point behaviours of the change of the photon orbit radius $\Delta \tilde{r}_{ps}$ and the minimum impact parameter $\Delta \tilde{u}_{ps}$ during the black hole phase transition for Hayward case. Red square dots are the numerical results and blue solid lines are our fitting results.}
\label{fitHayward}
\end{figure*}

This implies that $\ln  \Delta \tilde{r}_{ps}$ and $ \ln \Delta \tilde{u}_{ps}$ are linearly varies with $\ln (1-\tilde{T})$. We numerically fit the curve by varying $\tilde{T}$ from $0.99$ to $0.9999$. The numerically obtained results along with the fitting results are shown in Fig. \ref{fitHayward}. The numerical study reveals that, in the vicinity of critical point,
\begin{equation}
\Delta \tilde{r}_{ps} = 3.42124 (1-\tilde{T})^{0.500003},
\end{equation}
and
\begin{equation}
\Delta \tilde{u}_{ps} = 1.15021 (1-\tilde{T})^{0.500003}.
\end{equation}

This behaviour, i.e. $\Delta \tilde{r}_{ps} \sim (1-\tilde{T})^{1/2}$ and $\Delta \tilde{u}_{ps} \sim (1-\tilde{T})^{1/2}$, suggests that $\Delta \tilde{r}_{ps}$ and $\Delta \tilde{u}_{ps}$ can serve as the order parameters to characterise the phase transition. This, once again confirms our earlier observation on the connection between the photon orbits and thermodynamic phase transitions. The result shows that the numerical error in the calculation is negligible.

\section{Regular Bardeen Black Hole}
\label{bardeensecion}
In this section, we establish the connection between thermodynamic phase transition and the null geodesic for the Bardeen case. The Bardeen solution of the black hole in AdS spacetime has the following form \citep{Fan:2016hvf},
\begin{equation}
ds^2=-f(r)dt^2+\frac{1}{f(r)}dr^2+r^2d\Omega ^2
\end{equation}
with
\begin{equation}
f(r)=1-\frac{2 M r^2}{\left(g^2+r^2\right)^{3/2}}+\frac{8}{3} \pi  P r^2.
\end{equation}
As before, the pressure $P$ is related to the cosmological constant $\Lambda$ as $P=-
\Lambda /8\pi$. The condition $f(r_+)=0$ yields the mass of the black hole as,
\begin{equation}
M=\frac{\left(g^2+r_+^2\right)^{3/2} \left(8 \pi  P r_+^2+3\right)}{6 r_+^2}.
\end{equation}
The Hawking temperature can be easily obtained as,
\begin{align}
T&=\frac{\kappa}{2\pi}=\left. \frac{f'(r)}{4\pi} \right|_{r=r_+}\nonumber \\
&=\frac{2 P r_+^3}{g^2+r_+^2}+\frac{r_+}{4 \pi  \left(g^2+r_+^2\right)}-\frac{g^2}{2 \pi  r_+ \left(g^2+r_+^2\right)}.
\label{bardeentemp}
\end{align}

The first law of thermodynamics has the same form as that of Hayward case,
\begin{equation}
dM=TdS+\Psi dQ_m+VdP+\Pi d \alpha,
\end{equation}
with the variables having same meaning. In fact this is the generic form of the first law in the extended phase space for black holes with nonlinear electric/magnetic charges, which can be derived using a covariant approach \citep{Zhang:2016ilt}. The entropy of the black hole is,
\begin{equation}
S=\int \frac{dM}{T}=-\frac{2 \pi  g^3 }{r_+}\, _2F_1\left(-\frac{3}{2},-\frac{1}{2};\frac{1}{2};-\frac{r_+^2}{g^2}\right),
\end{equation}
where $\, _2F_1$ is the Hyper-geometric function. The volume $V$ can be obtained as,
\begin{equation}
    V=\left( \frac{\partial M}{\partial P}\right)_{S,Q_m,\alpha}=\frac{4}{3} \pi  \left(g^2+r_+^2\right)^{3/2}.
\end{equation}
Inverting the expression of Hawking temperature (Eq. \ref{bardeentemp}) for pressure we have the equation of state,
\begin{equation}
P=\frac{g^2}{4 \pi  r_+^4}+\frac{g^2 T}{2 r_+^3}-\frac{1}{8 \pi  r_+^2}+\frac{T}{2 r_+},
\end{equation}
from which an oscillatory behaviour of isotherms below critical temperature and hence the phase transition is evident, which is well studied \citep{Tzikas:2018cvs}. The critical values of the thermodynamic variables for this phase transition are determined, which are,

\begin{figure*}[t]
\centering
\subfigure[ref1][\, $\tilde{T}$ vs. $\tilde{r}_{ps}$]{\includegraphics[scale=0.9]{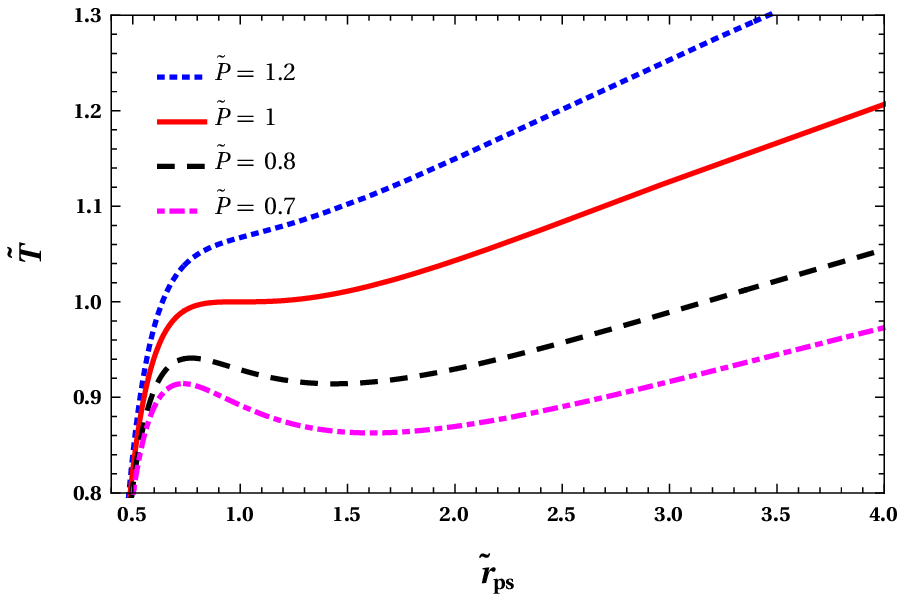}\label{TrBardeen}}
\qquad
\subfigure[ref2][\, $\tilde{T}$ vs. $\tilde{u}_{ps}$]{\includegraphics[scale=0.9]{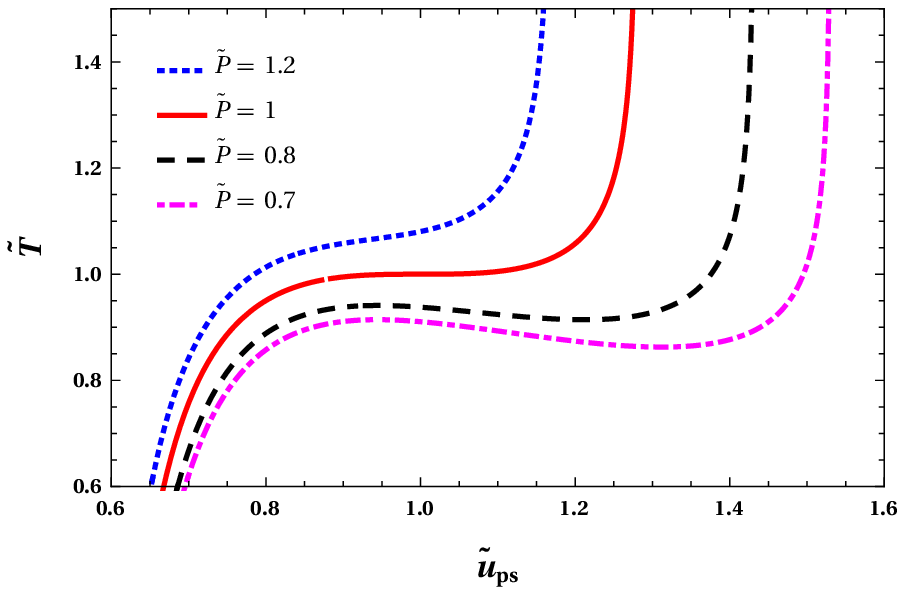}\label{TuBardeen}}
\caption{The behaviour of photon sphere radius $r_{ps}$ and minimum impact parameter $u_{ps}$ with temperature in reduced space for Bardeen case. These plots are for a fixed value of $g=0.8$ and the reduced pressure $\tilde{P}=0.7, 0.8, 1, 1.2$}
\label{TruBardeen}
\end{figure*}

\begin{figure*}[t]
\centering
\subfigure[ref1][\, $\tilde{P}$ vs. $\tilde{r}_{ps}$]{\includegraphics[scale=0.9]{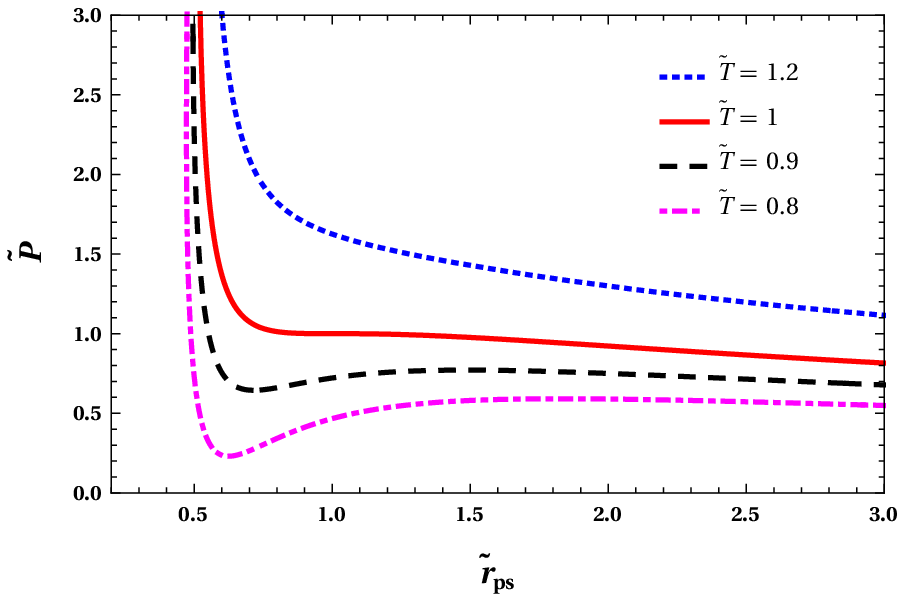}\label{PrBardeen}}
\qquad
\subfigure[ref2][\, $\tilde{P}$ vs. $\tilde{u}_{ps}$]{\includegraphics[scale=0.9]{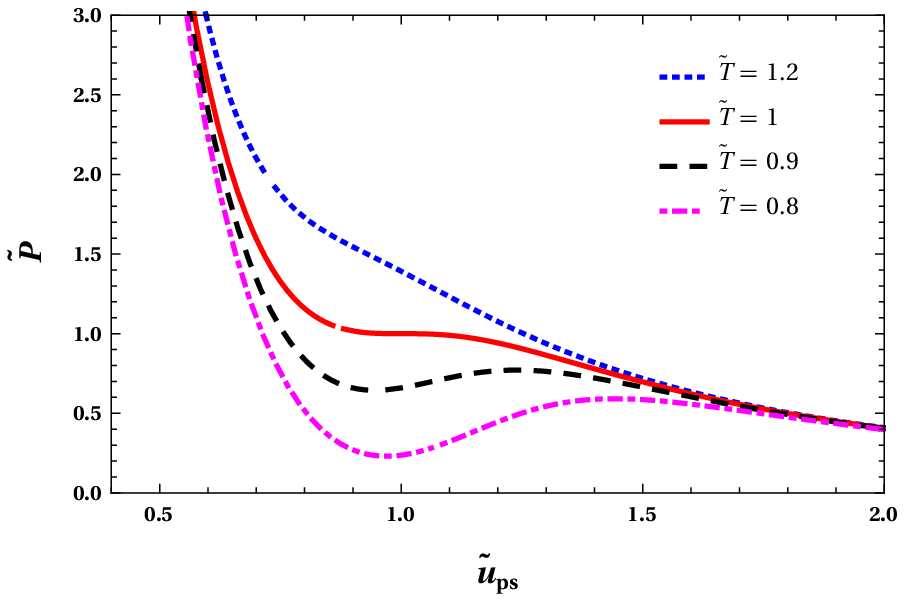}\label{PuBardeen}}
\caption{The behaviour of photon sphere radius $r_{ps}$ and minimum impact parameter $u_{ps}$ with pressure in reduced space for Bardeen case. These plots are for a fixed value of $g=0.8$ and the reduced temperature $\tilde{T}=0.8, 0.9, 1, 1.2$}
\label{PruBardeen}
\end{figure*}

\begin{equation}
T_{cB}=-\frac{\left(\sqrt{273}-17\right) \sqrt{\frac{1}{2} \left(\sqrt{273}+15\right)}}{24 \pi  g},
\end{equation}
\begin{equation}
P_{cB}=\frac{\sqrt{273}+27}{12 \left(\sqrt{273}+15\right)^2 \pi  g^2},
\end{equation}
\begin{equation}
S_{cB}=\frac{1}{2} \left(\sqrt{273}+15\right) \pi  g^2.
\end{equation}
As earlier, the critical values depend on the parameter $g$. The reduced thermodynamics variables can be defined from these critical variables as before. The study of coexistent physics is not trivial compared to Hayward black hole case as the analytical expression for the coexistence curve is not feasible. Therefore we investigate the critical behaviour of the black hole numerically in the next section. The coexistent curve which separates the SBH and LBH phases of the black hole can be obtained numerically by using the swallow tail behaviour of the Gibbs free energy, $G=M-TS$.

\subsection{Photon Orbit and Phase transition}

\begin{figure*}[t]
\centering
\subfigure[][\, $\tilde{r}_{ps}$ vs. $\tilde{T}$]{\includegraphics[scale=0.9]{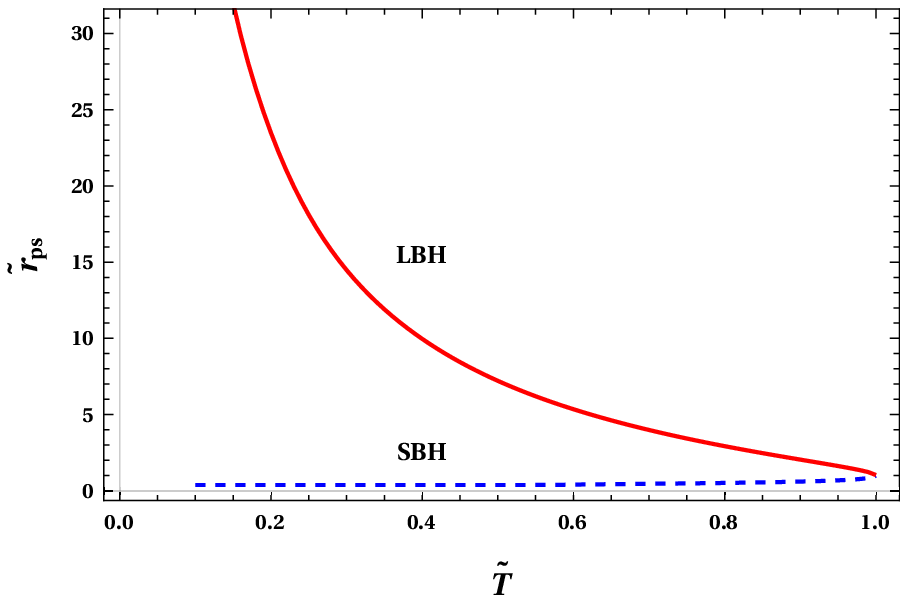}\label{rBardeen}}
\qquad
\subfigure[][\, $\tilde{u}_{ps}$ vs. $\tilde{T}$]{\includegraphics[scale=0.9]{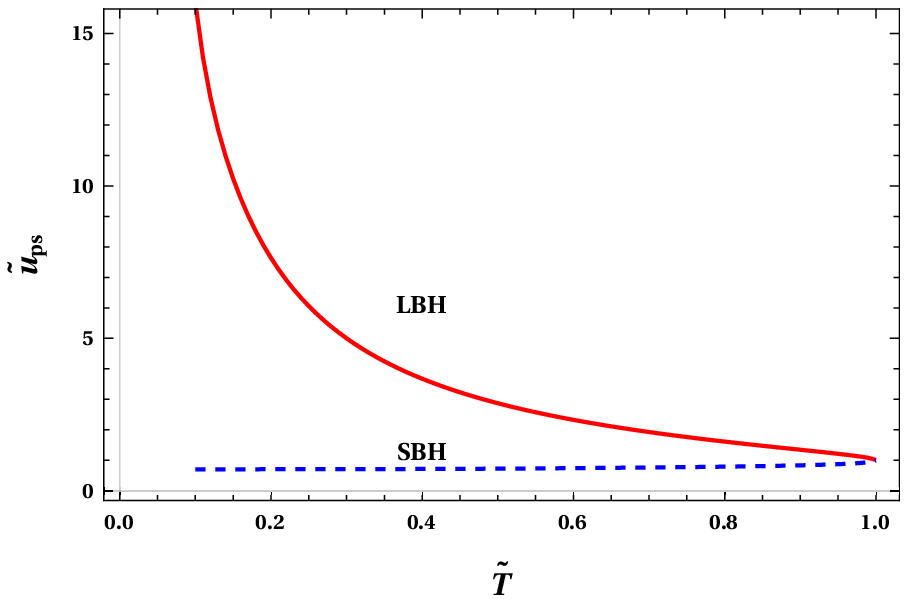}\label{uBardeen}}

\caption{The behaviour of photon sphere radius $r_{ps}$ and minimum impact parameter $u_{ps}$ of the photon orbit along the coexistence curve for Bardeen case}
\label{coexB}
\end{figure*}

\begin{figure*}[t]
\centering

\subfigure[][\,  $\Delta \tilde{r}_{ps}$ vs. $\tilde{T}$]{\includegraphics[scale=0.9]{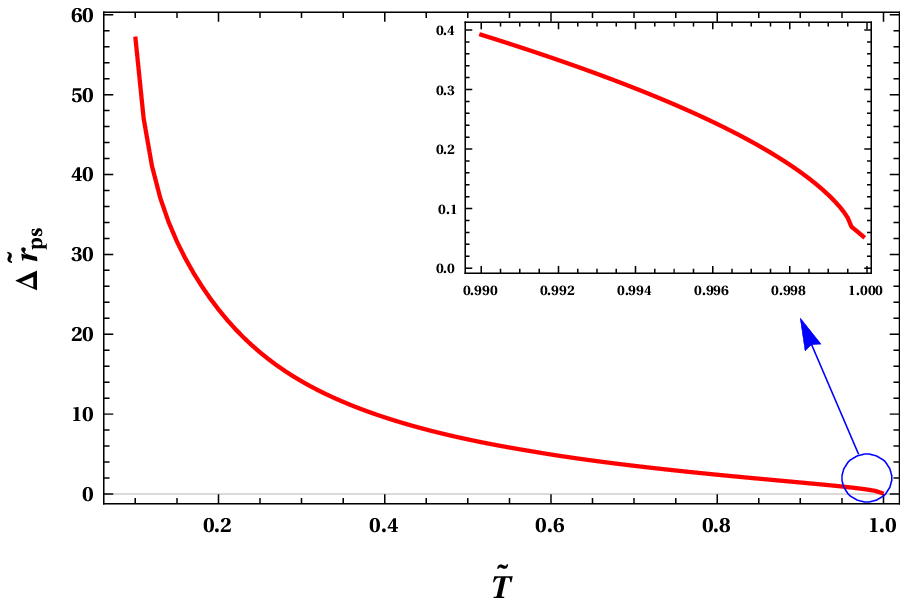}\label{drBardeen}}
\quad
\subfigure[][\, $\Delta \tilde{u}_{ps}$ vs. $\tilde{T}$]{\includegraphics[scale=0.9]{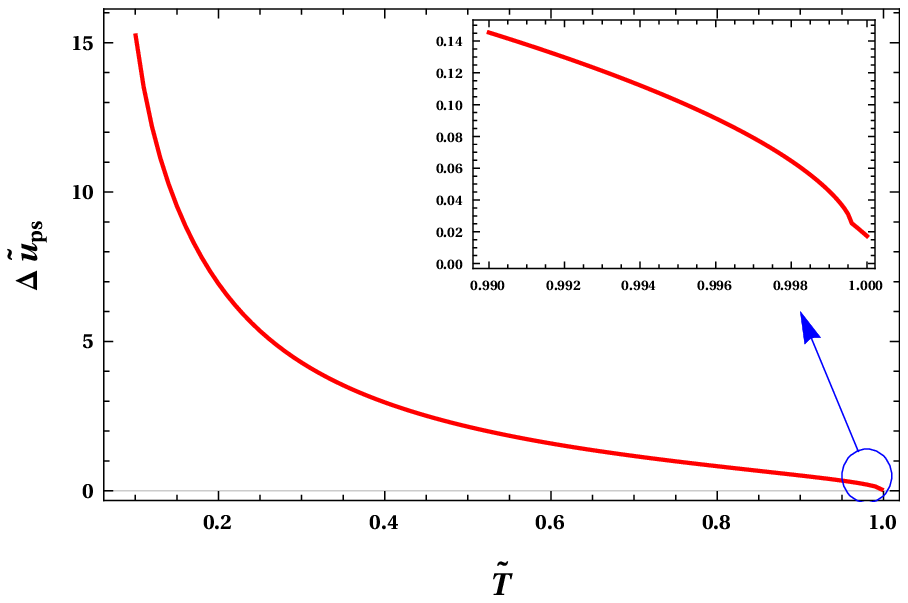}\label{duBardeen}}
\caption{ The behaviour of difference of the radii of the circular orbit $\Delta \tilde{r}_{ps}$ and the difference of the minimum impact parameter $\Delta \tilde{u}_{ps}$ along the coexistence curve for Bardeen case. The change of concavity near the critical point is shown in the inlets}
\label{diffcoexB}
\end{figure*}

The geodesic for the photon moving in the equatorial plane of regular Bardeen AdS black hole is analysed in the same line as in Hayward case. The behaviour of $\tilde{V}_{eff}=V_{eff}/E$ is similar to that of Hayward case. There exists a critical impact parameter $u_c$ which defines the unstable photon orbit. The photon approaching the black hole with impact parameter $u>u_c$ will be scattered and photon with $u<u_c$ will be absorbed by the black hole. However, quantitatively there is a  difference in the effective potential, for example height of the potential barrier is different in Bardeen and Hayward case for a given value of $g$.

The expression for the photon orbit radius is obtained by solving the second relation in Eq. (\ref{aneqn}) for the Bardeen background, which has a relatively simple form,
\begin{equation}
r_{ps}=\frac{2^{2/3} M^2}{\sqrt[3]{Z}}+M+\frac{\sqrt[3]{Z}}{2^{2/3}} ,
\label{rpsB}
\end{equation}
where,
\begin{equation}
Z=-15 g^2 M+\sqrt{15} \sqrt{15 g^4 M^2-8 g^2 M^4}+4 M^3.
\end{equation}

With the use of Eq. (\ref{rpsB}) and Eq. (\ref{upsequation}) we obtain the minimum impact parameter $u_{ps}$ for the Bardeen case. As argued earlier, the photon sphere radius and minimum impact parameter are the key quantities in probing the phase transition of the black hole. The isobars in $\tilde{T}$ vs. $\tilde{r}_{ps}$ and $\tilde{T}$ vs. $\tilde{u}_{ps}$ and isotherms in  $\tilde{P}$ vs. $\tilde{r}_{ps}$ and  $\tilde{P}$ vs. $\tilde{u}_{ps}$ planes show the corresponding vdW like phase transition in regular Bardeen black hole (Fig. \ref{TruBardeen} and Fig. \ref{PruBardeen}).

Unlike Hayward black hole case, it is not possible to solve the equal area law analytically for the isobars of the Bardeen black hole case. We have performed a numerical calculation to obtain the result. Using the result we studied the behaviour of photon orbit radius $r_{ps}$ and minimum impact parameter $u_{ps}$ along the coexistence line (Fig. \ref{coexB}). Here too, as in Hayward case, there is a sudden change for $r_{ps}$ and $u_{ps}$ during the phase transition. The changes $\Delta r_{ps}$ and $\Delta u_{ps}$ along the coexistence line is also studied numerically (Fig. \ref{diffcoexB}). The behaviour of the curve is as before. Using the curve fit formula of the form Eq. (\ref{logfit}), the numerical fit yields,

\begin{figure*}[t]
\centering
\subfigure[][\,  $\ln \Delta \tilde{r}_{ps}$ vs.  $ \ln (1- \tilde{T})$]{\includegraphics[scale=0.9]{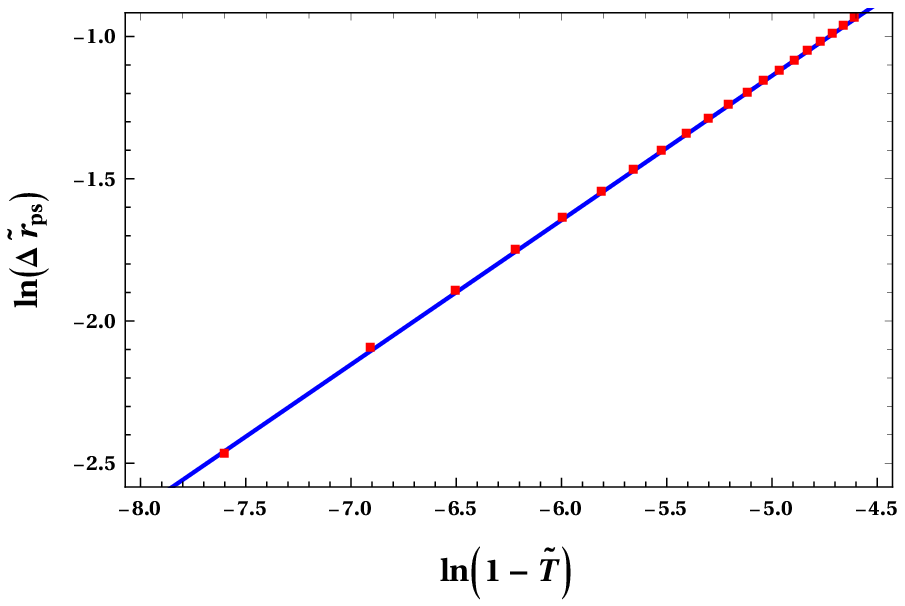}\label{drfitBardeen}}
\quad
\subfigure[][\, $\ln  \Delta \tilde{u}_{ps}$ vs. $ \ln (1-\tilde{T})$]{\includegraphics[scale=0.9]{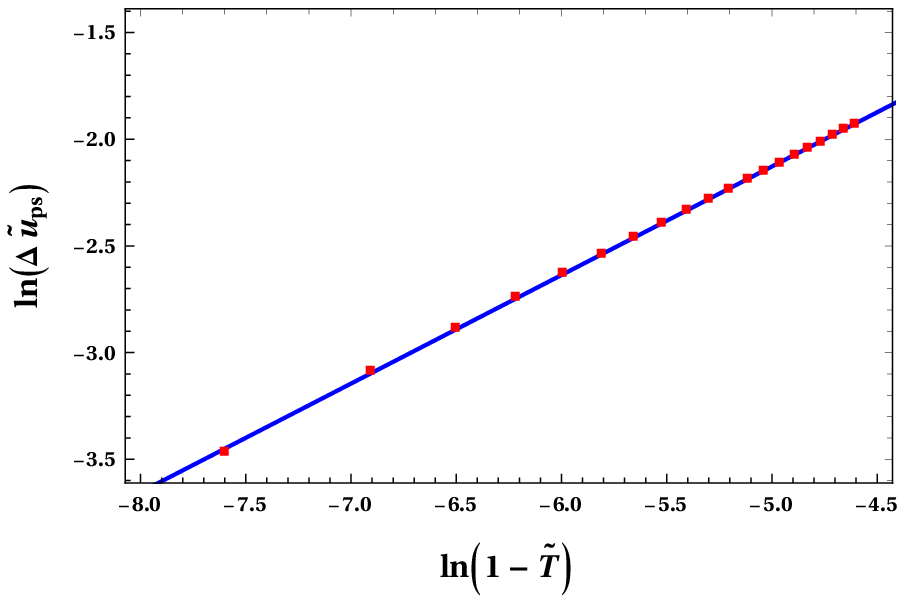}\label{dufitBardeen}}
\caption{Near critical point behaviours of the change of the photon orbit radius $\Delta \tilde{r}_{ps}$ and the minimum impact parameter $\Delta \tilde{u}_{ps}$ during the black hole phase transition for the Bardeen case. Red square dots are the numerical results and blue solid lines are our fitting results}
\label{fitBardeen}
\end{figure*}

\begin{equation}
\Delta \tilde{r}_{ps} = 4.04741 (1-\tilde{T})^{0.507242} ,
\end{equation}
and\begin{equation}
\Delta \tilde{u}_{ps} = 1.51445 (1-\tilde{T})^{0.508615}.
\end{equation}

This shows that $\Delta r_{ps}$ and $\Delta u_{ps}$ serve as order parameter of the phase transition with  critical exponent $1/2$. The numerical data and fitting results are shown in Fig. \ref{fitBardeen}.  In this case the numerical errors are within $1.72 \%$. Unlike Hayward case a finite tiny error present here as the coexistence curve also obtained numerically. From the connection between null geodesics and thermodynamic phase transition of Bardeen and Hayward cases, we can ascertain that the result is apparent for any regular spacetime background.

\section{Concluding Remarks}
\label{conclusion}
One of the most intriguing aspect of gravitational theories is the presence of physical singularity at the centre of black holes. Among the several ways of avoiding the black hole singularity, the non linear electrodynamics (NED) coupled to general relativity is widely discussed. One of the simple NED model which gives well defined magnetically charged black holes is the generic regular black holes \citep{Fan:2016hvf}. The interesting sub classes of this model are the Hayward and Bardeen black holes. Naturally, it is compelling to study the physical properties associated with the regular generic black holes and compare the differences and similarities to the generic black holes which exhibit a physical singularity.

In this article, using the formalism of unstable circular null geodesics for a class of regular black holes including Hayward-AdS and Bardeen-AdS spacetimes, we find a close connection between the gravity and thermodynamics in the extended phase space. The well-known van der Waals-like phase structure is probed via the photon orbit radius $r_{ps}$ and minimum critical impact parameter $u_{ps}$. In the reduced parameter space, the isobars and isotherms in each of these key parameters' plots show oscillatory behaviour below the critical values of the temperature $\tilde{T}$ and the pressure $\tilde{P}$, respectively. Such behaviours are in accordance with the van der Waals-like phase transition of the black holes. The disappearance of the first-order phase transition above the critical point is clearly seen in the isobar and isotherm plots of $r_{ps}$ and $u_{ps}$. Moreover, the differences $\Delta r_{ps}$ and $\Delta u_{ps}$ serve as order parameters for the critical behaviour. Furthermore, near the second-order phase transition points, these differences exhibit a change of concavity with critical exponents $\delta=1/2$. We studied the behaviour of $\Delta r_{ps}$ and $\Delta u_{ps}$ along the coexistence curve analytically for Hayward-AdS black hole, whereas, due to the difficulty in solving Bardeen case we adopted numerical method. The behaviour of $\Delta r_{ps}$ and $\Delta u_{ps}$ near the critical point are probed numerically in both the cases, with a negligible error in Hayward case and a mere $1.72\%$ in Bardeen case. In both cases we are lead to the same inference, from which we expect that the generic regular black holes are characterised by the same observed feature.  

Our results show that regular black holes are in close proximity with charged AdS black holes, in phase transition perspectives. Thus a regular modification to the electrovacuum solutions of Einstein field equations could be a possible candidate for probing its thermal properties. The thermal properties connecting the photon orbits may be useful to distinguish the regular black holes from Kerr one in AdS spacetimes and to test whether or not the regular black hole candidates are the black holes predicted by Einstein's relativity. This is one more interesting peripheral aspect of circular photon orbit which can disclose the observational signature of the thermodynamic phase transition. Our speculation is on the ground that photon orbit has strong astrophysical interest, and its ability to reflect the black hole phase transition.

\acknowledgments
The authors N.K.A. and A.R.C.L. thank the IIT Ropar for hospitality where part of this work was done during a visit. N.K.A. and A.R.C.L. would also like to thank U.G.C. Govt. of India for financial assistance under UGC-NET-SRF scheme. M.S.A.'s research is supported by the ISIRD grant 9-252/2016/IITRPR/708.

\bibliography{BibTex}

\providecommand{\href}[2]{#2}\begingroup\raggedright\begin{thebibliography}{10}

\bibitem{Hawking:1974sw}
S.~Hawking, \emph{{Particle Creation by Black Holes}},
  \href{https://doi.org/10.1007/BF02345020}{\emph{Commun. Math. Phys.}
  {\bfseries 43} (1975) 199}.

\bibitem{Bekenstein1973}
J.D.~Bekenstein, \emph{{Black holes and entropy}},
  \href{https://doi.org/10.1103/PhysRevD.7.2333}{\emph{Phys. Rev. D} {\bfseries
  7} (1973) 2333}.

\bibitem{Hawking1983}
S.W.~Hawking and D.N.~Page, \emph{{Thermodynamics of Black Holes in anti-De
  Sitter Space}}, \href{https://doi.org/10.1007/BF01208266}{\emph{Commun. Math.
  Phys.} {\bfseries 87} (1983) 577}.

\bibitem{Kastor:2009wy}
D.~Kastor, S.~Ray and J.~Traschen, \emph{{Enthalpy and the Mechanics of AdS
  Black Holes}},
  \href{https://doi.org/10.1088/0264-9381/26/19/195011}{\emph{Class. Quant.
  Grav.} {\bfseries 26} (2009) 195011}
  [\href{https://arxiv.org/abs/0904.2765}{{\ttfamily 0904.2765}}].

\bibitem{Dolan:2011xt}
B.P.~Dolan, \emph{{Pressure and volume in the first law of black hole
  thermodynamics}},
  \href{https://doi.org/10.1088/0264-9381/28/23/235017}{\emph{Class. Quant.
  Grav.} {\bfseries 28} (2011) 235017}
  [\href{https://arxiv.org/abs/1106.6260}{{\ttfamily 1106.6260}}].

\bibitem{Kubiznak2012}
D.~Kubiznak and R.B.~Mann, \emph{{P-V criticality of charged AdS black holes}},
  \href{https://doi.org/10.1007/JHEP07(2012)033}{\emph{JHEP} {\bfseries 07}
  (2012) 033} [\href{https://arxiv.org/abs/1205.0559}{{\ttfamily 1205.0559}}].

\bibitem{Gunasekaran2012}
S.~Gunasekaran, R.B.~Mann and D.~Kubiznak, \emph{{Extended phase space
  thermodynamics for charged and rotating black holes and Born-Infeld vacuum
  polarization}}, \href{https://doi.org/10.1007/JHEP11(2012)110}{\emph{JHEP}
  {\bfseries 11} (2012) 110} [\href{https://arxiv.org/abs/1208.6251}{{\ttfamily
  1208.6251}}].

\bibitem{Kubiznak:2016qmn}
D.~Kubiznak, R.B.~Mann and M.~Teo, \emph{{Black hole chemistry: thermodynamics
  with Lambda}}, \href{https://doi.org/10.1088/1361-6382/aa5c69}{\emph{Class.
  Quant. Grav.} {\bfseries 34} (2017) 063001}
  [\href{https://arxiv.org/abs/1608.06147}{{\ttfamily 1608.06147}}].

\bibitem{Cardoso:2008bp}
V.~Cardoso, A.S.~Miranda, E.~Berti, H.~Witek and V.T.~Zanchin, \emph{{Geodesic
  stability, Lyapunov exponents and quasinormal modes}},
  \href{https://doi.org/10.1103/PhysRevD.79.064016}{\emph{Phys. Rev. D}
  {\bfseries 79} (2009) 064016}
  [\href{https://arxiv.org/abs/0812.1806}{{\ttfamily 0812.1806}}].

\bibitem{Stefanov:2010xz}
I.Z.~Stefanov, S.S.~Yazadjiev and G.G.~Gyulchev, \emph{{Connection between
  Black-Hole Quasinormal Modes and Lensing in the Strong Deflection Limit}},
  \href{https://doi.org/10.1103/PhysRevLett.104.251103}{\emph{Phys. Rev. Lett.}
  {\bfseries 104} (2010) 251103}
  [\href{https://arxiv.org/abs/1003.1609}{{\ttfamily 1003.1609}}].

\bibitem{Liu:2014gvf}
Y.~Liu, D.-C.~Zou and B.~Wang, \emph{{Signature of the Van der Waals like
  small-large charged AdS black hole phase transition in quasinormal modes}},
  \href{https://doi.org/10.1007/JHEP09(2014)179}{\emph{JHEP} {\bfseries 09}
  (2014) 179} [\href{https://arxiv.org/abs/1405.2644}{{\ttfamily 1405.2644}}].

\bibitem{Wei:2017mwc}
S.-W.~Wei and Y.-X.~Liu, \emph{{Photon orbits and thermodynamic phase
  transition of $d$-dimensional charged AdS black holes}},
  \href{https://doi.org/10.1103/PhysRevD.97.104027}{\emph{Phys. Rev. D}
  {\bfseries 97} (2018) 104027}
  [\href{https://arxiv.org/abs/1711.01522}{{\ttfamily 1711.01522}}].

\bibitem{Wei:2018aqm}
S.-W.~Wei, Y.-X.~Liu and Y.-Q.~Wang, \emph{{Probing the relationship between
  the null geodesics and thermodynamic phase transition for rotating Kerr-AdS
  black holes}}, \href{https://doi.org/10.1103/PhysRevD.99.044013}{\emph{Phys.
  Rev. D} {\bfseries 99} (2019) 044013}
  [\href{https://arxiv.org/abs/1807.03455}{{\ttfamily 1807.03455}}].

\bibitem{Xu:2019yub}
Y.-M.~Xu, H.-M.~Wang, Y.-X.~Liu and S.-W.~Wei, \emph{{Photon sphere and
  reentrant phase transition of charged Born-Infeld-AdS black holes}},
  \href{https://doi.org/10.1103/PhysRevD.100.104044}{\emph{Phys. Rev. D}
  {\bfseries 100} (2019) 104044}
  [\href{https://arxiv.org/abs/1906.03334}{{\ttfamily 1906.03334}}].

\bibitem{Chabab:2019kfs}
M.~Chabab, H.~El~Moumni, S.~Iraoui and K.~Masmar, \emph{{Probing correlation
  between photon orbits and phase structure of charged AdS black hole in
  massive gravity background}},
  \href{https://doi.org/10.1142/S0217751X19502312}{\emph{Int. J. Mod. Phys. A}
  {\bfseries 34} (2020) 1950231}
  [\href{https://arxiv.org/abs/1902.00557}{{\ttfamily 1902.00557}}].

\bibitem{Li:2019dai}
H.~Li, Y.~Chen and S.-J.~Zhang, \emph{{Photon orbits and phase transitions in
  Born-Infeld-dilaton black holes}},
  \href{https://doi.org/10.1016/j.nuclphysb.2020.114975}{\emph{Nucl. Phys. B}
  {\bfseries 954} (2020) 114975}
  [\href{https://arxiv.org/abs/1908.09570}{{\ttfamily 1908.09570}}].

\bibitem{Han:2018ooi}
S.-Z.~Han, J.~Jiang, M.~Zhang and W.-B.~Liu, \emph{{Photon orbits and
  thermodynamic phase transition in Gauss-Bonnet AdS black holes}},
  {\emph{arXiv:1812.11862 [gr-qc]} (2018) }
  [\href{https://arxiv.org/abs/1812.11862}{{\ttfamily 1812.11862}}].

\bibitem{Hegde:2020yrd}
K.~Hegde, A.~Naveena~Kumara, C.A.~Rizwan, M.S.~Ali and A.K.~M, \emph{{Null
  Geodesics and Thermodynamic Phase Transition of Four-Dimensional Gauss-Bonnet
  AdS Black Hole}},  \href{https://arxiv.org/abs/2007.10259}{{\ttfamily
  2007.10259}}.

\bibitem{Zhang:2019tzi}
M.~Zhang, S.-Z.~Han, J.~Jiang and W.-B.~Liu, \emph{{Circular orbit of a test
  particle and phase transition of a black hole}},
  \href{https://doi.org/10.1103/PhysRevD.99.065016}{\emph{Phys. Rev. D}
  {\bfseries 99} (2019) 065016}
  [\href{https://arxiv.org/abs/1903.08293}{{\ttfamily 1903.08293}}].

\bibitem{Bhamidipati:2018yqy}
B.~Chandrasekhar and S.~Mohapatra, \emph{{A Note on Circular Geodesics and
  Phase Transitions of Black Holes}},
  \href{https://doi.org/10.1016/j.physletb.2019.02.042}{\emph{Phys. Lett. B}
  {\bfseries 791} (2019) 367}
  [\href{https://arxiv.org/abs/1805.05088}{{\ttfamily 1805.05088}}].

\bibitem{Wei:2019jve}
S.-W.~Wei and Y.-X.~Liu, \emph{{Null Geodesics, Quasinormal Modes, and
  Thermodynamic Phase Transition for Charged Black Holes in Asymptotically Flat
  and dS Spacetimes}}, {\emph{arXiv:1909.11911 [gr-qc]} (2019) }
  [\href{https://arxiv.org/abs/1909.11911}{{\ttfamily 1909.11911}}].

\bibitem{Hawking:1969sw}
S.W.~Hawking and R.~Penrose, \emph{{The Singularities of gravitational collapse
  and cosmology}}, \href{https://doi.org/10.1098/rspa.1970.0021}{\emph{Proc.
  Roy. Soc. Lond. A} {\bfseries 314} (1970) 529}.

\bibitem{Hawking:1973uf}
S.W.~Hawking and G.F.R.~Ellis, \emph{{The Large Scale Structure of
  Space-Time}}, Cambridge Monographs on Mathematical Physics, Cambridge
  University Press (2011),
  \href{https://doi.org/10.1017/CBO9780511524646}{10.1017/CBO9780511524646}.

\bibitem{1966JETP...22..241S}
A.D.~{Sakharov}, \emph{{The Initial Stage of an Expanding Universe and the
  Appearance of a Nonuniform Distribution of Matter}}, {\emph{Soviet Journal of
  Experimental and Theoretical Physics} {\bfseries 22} (1966) 241}.

\bibitem{1966JETP...22..378G}
E.B.~{Gliner}, \emph{{Algebraic Properties of the Energy-momentum Tensor and
  Vacuum-like States o$^{+}$ Matter}}, {\emph{Soviet Journal of Experimental
  and Theoretical Physics} {\bfseries 22} (1966) 378}.

\bibitem{bardeen1968non}
J.~Bardeen, \emph{Non-singular general-relativistic gravitational collapse, in
  proceedings of the international conference gr5}, {\emph{Tbilisi, USSR}
  (1968) 174}.

\bibitem{Hayward:2005gi}
S.A.~Hayward, \emph{{Formation and evaporation of regular black holes}},
  \href{https://doi.org/10.1103/PhysRevLett.96.031103}{\emph{Phys. Rev. Lett.}
  {\bfseries 96} (2006) 031103}
  [\href{https://arxiv.org/abs/gr-qc/0506126}{{\ttfamily gr-qc/0506126}}].

\bibitem{AyonBeato:1998ub}
E.~Ayon-Beato and A.~Garcia, \emph{{Regular black hole in general relativity
  coupled to nonlinear electrodynamics}},
  \href{https://doi.org/10.1103/PhysRevLett.80.5056}{\emph{Phys. Rev. Lett.}
  {\bfseries 80} (1998) 5056}
  [\href{https://arxiv.org/abs/gr-qc/9911046}{{\ttfamily gr-qc/9911046}}].

\bibitem{AyonBeato:2000zs}
E.~Ayon-Beato and A.~Garcia, \emph{{The Bardeen model as a nonlinear magnetic
  monopole}}, \href{https://doi.org/10.1016/S0370-2693(00)01125-4}{\emph{Phys.
  Lett. B} {\bfseries 493} (2000) 149}
  [\href{https://arxiv.org/abs/gr-qc/0009077}{{\ttfamily gr-qc/0009077}}].

\bibitem{Man:2013hpa}
J.~Man and H.~Cheng, \emph{{The calculation of the thermodynamic quantities of
  the Bardeen black hole}},
  \href{https://doi.org/10.1007/s10714-013-1660-4}{\emph{Gen. Rel. Grav.}
  {\bfseries 46} (2014) 1660}
  [\href{https://arxiv.org/abs/1304.5686}{{\ttfamily 1304.5686}}].

\bibitem{Man:2013hza}
J.~Man and H.~Cheng, \emph{{The description of phase transition of Bardeen
  black hole in the Ehrenfest scheme}}, {\emph{arXiv:1312.6566 [hep-th]} (2013)
  } [\href{https://arxiv.org/abs/1312.6566}{{\ttfamily 1312.6566}}].

\bibitem{Abdujabbarov:2016hnw}
A.~Abdujabbarov, M.~Amir, B.~Ahmedov and S.G.~Ghosh, \emph{{Shadow of rotating
  regular black holes}},
  \href{https://doi.org/10.1103/PhysRevD.93.104004}{\emph{Phys. Rev. D}
  {\bfseries 93} (2016) 104004}
  [\href{https://arxiv.org/abs/1604.03809}{{\ttfamily 1604.03809}}].

\bibitem{Amir:2016cen}
M.~Amir and S.G.~Ghosh, \emph{{Shapes of rotating nonsingular black hole
  shadows}}, \href{https://doi.org/10.1103/PhysRevD.94.024054}{\emph{Phys. Rev.
  D} {\bfseries 94} (2016) 024054}
  [\href{https://arxiv.org/abs/1603.06382}{{\ttfamily 1603.06382}}].

\bibitem{Flachi:2012nv}
A.~Flachi and J.P.S.~Lemos, \emph{{Quasinormal modes of regular black holes}},
  \href{https://doi.org/10.1103/PhysRevD.87.024034}{\emph{Phys. Rev. D}
  {\bfseries 87} (2013) 024034}
  [\href{https://arxiv.org/abs/1211.6212}{{\ttfamily 1211.6212}}].

\bibitem{Eiroa:2010wm}
E.F.~Eiroa and C.M.~Sendra, \emph{{Gravitational lensing by a regular black
  hole}}, \href{https://doi.org/10.1088/0264-9381/28/8/085008}{\emph{Class.
  Quant. Grav.} {\bfseries 28} (2011) 085008}
  [\href{https://arxiv.org/abs/1011.2455}{{\ttfamily 1011.2455}}].

\bibitem{Aros:2019quj}
R.~Aros and M.~Estrada, \emph{{Regular black holes and its thermodynamics in
  Lovelock gravity}},
  \href{https://doi.org/10.1140/epjc/s10052-019-6783-7}{\emph{Eur. Phys. J. C}
  {\bfseries 79} (2019) 259}
  [\href{https://arxiv.org/abs/1901.08724}{{\ttfamily 1901.08724}}].

\bibitem{Nam:2018ltb}
C.H.~Nam, \emph{{Non-linear charged AdS black hole in massive gravity}},
  \href{https://doi.org/10.1140/epjc/s10052-018-6498-1}{\emph{Eur. Phys. J. C}
  {\bfseries 78} (2018) 1016}.

\bibitem{PhysRevD.98.084025}
M.S.~Ali and S.G.~Ghosh, \emph{Exact $d$-dimensional bardeen-de sitter black
  holes and thermodynamics},
  \href{https://doi.org/10.1103/PhysRevD.98.084025}{\emph{Phys. Rev. D}
  {\bfseries 98} (2018) 084025}.

\bibitem{Kumar:2018vsm}
A.~Kumar, D.~Veer~Singh and S.G.~Ghosh, \emph{{$D$-dimensional Bardeen-AdS
  black holes in Einstein-Gauss-Bonnet theory}},
  \href{https://doi.org/10.1140/epjc/s10052-019-6773-9}{\emph{Eur. Phys. J. C}
  {\bfseries 79} (2019) 275}
  [\href{https://arxiv.org/abs/1808.06498}{{\ttfamily 1808.06498}}].

\bibitem{Kumara:2020mvo}
A.~Naveena~Kumara, C.A.~Rizwan, K.~Hegde, A.K.~M. and M.S.~Ali,
  \emph{{Microstructure and continuous phase transition of a regular Hayward
  black hole in anti-de Sitter spacetime}}, {\emph{arXiv:2003.00889 [gr-qc]}
  (2020) } [\href{https://arxiv.org/abs/2003.00889}{{\ttfamily 2003.00889}}].

\bibitem{Kumara:2020ucr}
A.~Naveena~Kumara, C.A.~Rizwan, K.~Hegde and A.K.~M, \emph{{Repulsive
  Interactions in the Microstructure of Regular Hayward Black Hole in Anti-de
  Sitter Spacetime}},
  \href{https://doi.org/10.1016/j.physletb.2020.135556}{\emph{Phys. Lett. B}
  {\bfseries 807} (2020) 135556}
  [\href{https://arxiv.org/abs/2003.10175}{{\ttfamily 2003.10175}}].

\bibitem{Fan:2016hvf}
Z.-Y.~Fan and X.~Wang, \emph{{Construction of Regular Black Holes in General
  Relativity}}, \href{https://doi.org/10.1103/PhysRevD.94.124027}{\emph{Phys.
  Rev. D} {\bfseries 94} (2016) 124027}
  [\href{https://arxiv.org/abs/1610.02636}{{\ttfamily 1610.02636}}].

\bibitem{Ma:2014qma}
M.-S.~Ma and R.~Zhao, \emph{{Corrected form of the first law of thermodynamics
  for regular black holes}},
  \href{https://doi.org/10.1088/0264-9381/31/24/245014}{\emph{Class. Quant.
  Grav.} {\bfseries 31} (2014) 245014}
  [\href{https://arxiv.org/abs/1411.0833}{{\ttfamily 1411.0833}}].

\bibitem{Fan:2016rih}
Z.-Y.~Fan, \emph{{Critical phenomena of regular black holes in anti-de Sitter
  space-time}},
  \href{https://doi.org/10.1140/epjc/s10052-017-4830-9}{\emph{Eur. Phys. J. C}
  {\bfseries 77} (2017) 266}
  [\href{https://arxiv.org/abs/1609.04489}{{\ttfamily 1609.04489}}].

\bibitem{Bozza:2002zj}
V.~Bozza, \emph{{Gravitational lensing in the strong field limit}},
  \href{https://doi.org/10.1103/PhysRevD.66.103001}{\emph{Phys. Rev. D}
  {\bfseries 66} (2002) 103001}
  [\href{https://arxiv.org/abs/gr-qc/0208075}{{\ttfamily gr-qc/0208075}}].

\bibitem{Zhang:2016ilt}
Y.~Zhang and S.~Gao, \emph{{First law and Smarr formula of black hole mechanics
  in nonlinear gauge theories}},
  \href{https://doi.org/10.1088/1361-6382/aac9d4}{\emph{Class. Quant. Grav.}
  {\bfseries 35} (2018) 145007}
  [\href{https://arxiv.org/abs/1610.01237}{{\ttfamily 1610.01237}}].

\bibitem{Tzikas:2018cvs}
A.G.~Tzikas, \emph{{Bardeen black hole chemistry}},
  \href{https://doi.org/10.1016/j.physletb.2018.11.036}{\emph{Phys. Lett. B}
  {\bfseries 788} (2019) 219}
  [\href{https://arxiv.org/abs/1811.01104}{{\ttfamily 1811.01104}}].

\end{thebibliography}\endgroup
  
\end{document}